\documentclass[10pt,twocolumn]{article}

\usepackage[margin=0.78in]{geometry}
\usepackage{times}
\usepackage{microtype}
\usepackage{graphicx}
\usepackage{booktabs}
\usepackage{amsmath}
\usepackage{array}
\usepackage{enumitem}
\usepackage{listings}
\usepackage{xcolor}
\usepackage{caption}
\usepackage[hidelinks]{hyperref}
\usepackage[round]{natbib}

\setlist[itemize]{leftmargin=*,noitemsep,topsep=2pt}
\setlist[enumerate]{leftmargin=*,noitemsep,topsep=2pt}
\title{A Systems-Level Analysis of Sensitivity, Robustness, and Stability in Retrieval-Augmented Generation}

\author{
  Bharath Simha Reddy Muthyam \\
  Independent Researcher \\
  \texttt{bharathreddymuthyam@gmail.com}
}

\date{}

\begin{document}
\maketitle

\begin{abstract}
Retrieval-Augmented Generation (RAG) systems are often evaluated using final answer accuracy, even though their failures can originate from preprocessing, retrieval, context packing, or generation. This paper presents a controlled empirical study of RAG sensitivity, robustness, and stability across 56 experimental runs. We evaluate how chunk size, retrieval depth (\texttt{top\_k}), embedding-based reranking, probabilistic retrieval noise, and repeated seeded runs affect retrieval, context packing, and generation behavior. Using a fixed 500-question QA subset mapped to 20,958 unique corpus contexts, we analyze both final answer metrics and intermediate failure modes. Across these experiments, retrieval-oriented metrics improved under broader retrieval settings, while downstream exact-match and F1 scores often behaved non-monotonically. We also observe preprocessing-induced answer loss under smaller chunk sizes, progressive degradation under retrieval corruption, and higher observed variance in broader retrieval regimes. These findings suggest that RAG evaluation should include sensitivity, robustness, stability, and multi-stage failure analysis rather than relying only on final answer accuracy.
\end{abstract}

\section{Introduction}

Retrieval-Augmented Generation (RAG) systems are often evaluated using final answer accuracy, despite being multi-stage pipelines whose failures can originate from preprocessing, retrieval, context packing, or generation. Instead of relying only on model parameters, a RAG system retrieves relevant passages from a corpus and conditions generation on the retrieved context. This design is attractive for knowledge-intensive question answering because the retrieval component can expose the generator to task-specific or up-to-date information.

However, RAG reliability is difficult to understand from final answer quality alone. A low exact-match or F1 score may result from several different causes: the correct source context may not have been retrieved, the answer-bearing evidence may have been retrieved but excluded during context packing, or the generator may fail even when the answer is present in the prompt. In addition, preprocessing choices such as chunk size and filtering can remove answer-bearing text before retrieval begins. These stages create a multi-step reliability problem rather than a single model-accuracy problem.

This paper studies RAG as an empirical system under controlled perturbations. We do not aim to introduce a new RAG architecture or optimize for a leaderboard result. Instead, we ask how the behavior of a fixed RAG pipeline changes when retrieval configuration, retrieval corruption, and random seed conditions are varied. This study intentionally prioritizes behavioral analysis over leaderboard optimization. The goal is to understand tradeoffs among retrieval coverage, robustness, latency, stability, and generation quality.

We organize the study around three research questions:

\begin{itemize}
  \item \textbf{RQ1.} How sensitive is RAG performance to retrieval configuration changes such as chunk size, \texttt{top\_k}, and reranking?
  \item \textbf{RQ2.} How robust is RAG under controlled retrieval corruption?
  \item \textbf{RQ3.} How stable is RAG across repeated seeded runs with identical configurations?
\end{itemize}

Across these questions, a recurring theme emerges: retrieval improvements do not necessarily produce generation improvements. Increasing retrieval depth improves Hit@k and often improves whether the gold answer appears in the packed context, but downstream EM and F1 can plateau or decline. This retrieval-generation mismatch indicates that RAG reliability depends on how retrieved evidence is converted into usable generation context, not only on whether relevant evidence is retrieved.

This paper makes the following contributions:
\begin{itemize}
  \item We present a controlled empirical evaluation framework for analyzing sensitivity, robustness, and stability in RAG systems.
  \item We quantify preprocessing-induced answer loss across chunk sizes and show how failure can arise before retrieval begins.
  \item We characterize retrieval-generation tradeoffs, including cases where retrieval coverage improves while generation quality declines.
  \item We evaluate robustness under controlled retrieval corruption using probabilistic chunk replacement.
  \item We measure seed-level stability across repeated identical configurations under a moderate noise condition.
  \item We decompose final answer errors into retrieval, packing, and generation failure modes.
\end{itemize}

\section{Related Work}

\subsection{Retrieval-Augmented Generation}

RAG systems combine information retrieval with neural generation to improve performance on knowledge-intensive tasks. \citet{lewis2020rag} introduced retrieval-augmented generation as a framework in which retrieved passages condition a sequence-to-sequence generator, making retrieval part of the answer-generation process. More recent RAG surveys and evaluation work describe a broader ecosystem of retrieval, augmentation, and generation components \citep{gao2023rag,saadfalcon2024ares}. These studies motivate RAG as a modular pipeline, but many evaluations still emphasize final answer quality or component-level scores.

The present study follows this retrieval-then-generation pattern, but focuses on behavioral analysis rather than model development. We study how a fixed pipeline responds to configuration changes, retrieval corruption, and repeated seeded runs, with particular attention to how failures propagate across stages.

\subsection{Sensitivity in Retrieval Configuration}

RAG behavior depends on configuration choices such as chunk size, retrieval depth, context budget, and reranking strategy. These choices affect both the retriever's chance of finding answer-bearing evidence and the generator's ability to use that evidence once it is packed into the prompt. Existing benchmark and evaluation work has examined retrieval-augmented performance under different retrieval and context conditions \citep{chen2023rgb,gao2023rag}, but sensitivity is often treated as a tuning concern rather than as a system behavior to analyze directly. Our sensitivity experiments make these configuration tradeoffs explicit by varying chunk size, \texttt{top\_k}, and reranking while holding the dataset and generator fixed.

\subsection{Dense Retrieval and Reranking}

Dense retrieval methods such as Dense Passage Retrieval (DPR) represent queries and passages in a shared vector space, enabling semantic retrieval through nearest-neighbor search \citep{karpukhin2020dpr}. Sentence-BERT and compact models such as MiniLM provide sentence-level embeddings that are commonly used in applied retrieval pipelines \citep{reimers2019sentencebert,wang2020minilm}. FAISS provides efficient similarity search over dense vector indexes \citep{johnson2021faiss}.

Recent reranking work has further emphasized that evidence selection should account for downstream generation behavior, including dynamic adjustment of retrieved-document order and count and boundary-aware evidence selection under noisy retrieval \citep{sun2025dynamicrag,sun2026barrag}.

This study uses MiniLM embeddings and FAISS indexing. Reranking is implemented using an embedding-based similarity signal. Under the tested conditions, reranking had limited measurable effect, potentially influenced by the similarity between the retrieval and reranking signals. We treat this as a pipeline-specific observation: reranking may help when it introduces a stronger or more diverse relevance signal, but reordering alone does not guarantee downstream generation gains.

\subsection{Robustness and Noise in RAG}

Recent RAG robustness work has shown that retrieval-augmented systems can be sensitive to irrelevant, noisy, or perturbed context \citep{chen2023rgb,querylevelragrobustness2025}. Robustness studies are important because deployed retrieval systems rarely return perfectly clean evidence. Query-level RAG robustness work is especially relevant to this paper because it treats RAG performance as a behavior that changes under perturbation rather than a single static score.

Our robustness experiment uses controlled random retrieval corruption. Each retrieved chunk is independently replaced by a random corpus chunk with probability \texttt{noise\_percent / 100}. This keeps retrieval depth fixed while isolating the effect of corrupted evidence.

\subsection{Context Utilization and Failure Propagation}

A growing body of work shows that language models do not always use long context reliably. The ``Lost in the Middle'' finding is particularly relevant: models can struggle to use relevant information depending on its location and surrounding context \citep{liu2023lost}. This supports the broader concern that retrieving more context may not monotonically improve answer quality.

Our results are consistent with this concern. We observe regimes where Hit@k and gold-in-packed-context increase while EM and F1 decline, suggesting that extra context can introduce attention dilution, irrelevant evidence, or competing signals. We also separate retrieval failure from packing and generation failure, which helps explain where final answer errors arise.

\subsection{Stability and Variance in Generation Systems}

RAG reproducibility depends not only on model weights and datasets, but also on retrieval randomness, noise injection, and pipeline-level stochasticity. Even when a configuration is fixed, repeated runs can vary if retrieval corruption or other seeded components change. This study treats stability as an empirical property of a RAG configuration and evaluates variance across repeated seeded runs.

\section{Methodology and Evaluation Framework}

\subsection{Pipeline Overview}

\begin{figure*}[t]
  \centering
  \includegraphics[width=0.98\textwidth]{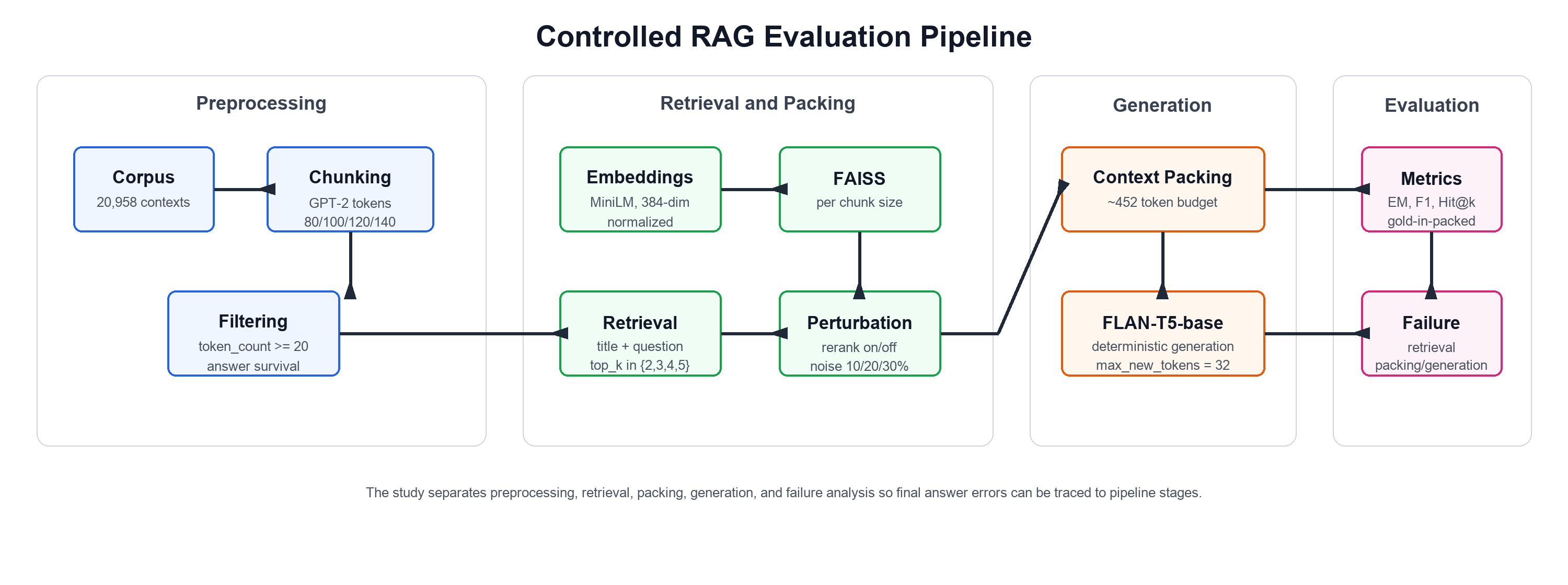}
  \caption{Pipeline architecture for the controlled RAG evaluation framework. The study separates preprocessing, retrieval, packing, generation, and failure analysis so final answer errors can be traced to pipeline stages.}
  \label{fig:pipeline}
\end{figure*}

Figure~\ref{fig:pipeline} shows the full pipeline. The pipeline separates preprocessing, retrieval, packing, generation, and failure analysis. This separation is central to the paper because final answer failure can originate from different stages.

\subsection{Dataset and Corpus}

The QA dataset used is SQuAD v1.1~\citep{rajpurkar2016squad}. We combined the train and validation splits, deduplicated examples by context paragraph, and constructed a corpus of 20,958 unique contexts. From the QA rows, we sampled a fixed 500-question evaluation subset using a fixed random seed. This subset was frozen across all 56 runs so that differences across configurations reflect pipeline behavior rather than changes in the evaluation questions.

Each unique context was assigned a constructed \texttt{context\_id}, and each QA pair was mapped back to the \texttt{context\_id} of its source paragraph. The \texttt{context\_id} is used as the retrieval correctness signal. This is important because answer-string matching alone is insufficient for retrieval evaluation: short answers such as ``South'', ``5,000'', or ``VH1'' may appear in many unrelated chunks. A retrieved chunk is therefore considered a retrieval hit when it comes from the gold \texttt{context\_id}, not merely when it contains a matching answer string.

The preprocessing pipeline was designed to preserve this alignment across corpus construction, chunking, filtering, retrieval, and evaluation. After chunking and filtering, the system checks whether each answer-bearing example still has recoverable evidence. This validation is a deliberate evaluation-stage check: answer-bearing text can be fragmented across chunks or removed by filtering before retrieval begins. We therefore track preprocessing-induced failures separately from downstream retrieval, packing, and generation failures. Under the smallest tested chunk size, this validation identified 23 of the 500 QA pairs as preprocessing-unanswerable---failures that no retriever or generator could recover. Treating preprocessing as an evaluated stage keeps the failure analysis grounded in the full RAG pipeline rather than only the final generation output.

\subsection{Chunking and Filtering}

Contexts are tokenized using the GPT-2 tokenizer and split into non-overlapping token chunks. We evaluate chunk sizes of 80, 100, 120, and 140 tokens. Non-overlapping chunking is used to isolate the effect of chunk size without introducing overlap-based redundancy.

Chunks with fewer than 20 tokens are filtered out:
\begin{equation}
\texttt{token\_count} \geq 20.
\end{equation}

For each chunk size, we record whether the gold answer survives preprocessing. This step captures preprocessing-induced answer loss before retrieval begins.

\subsection{Embedding and Indexing}

Each chunk-size setting has its own embedding set and FAISS index. We use \texttt{sentence-transformers/all-MiniLM-L6-v2}, which produces 384-dimensional embeddings. Embeddings are normalized before indexing. Separate indexes are necessary because each chunk size produces a different chunk universe and different metadata alignment.

\subsection{Retrieval and Reranking}

The retrieval query concatenates the title and question:
\begin{equation}
q = \text{title} + \text{question}.
\end{equation}

The title provides topic grounding for context-dependent questions, while the question provides the local information need. FAISS retrieves the top \(k\) chunks, with:
\begin{equation}
\texttt{top\_k} \in \{2,3,4,5\}.
\end{equation}

Reranking is evaluated in two states: on and off. When enabled, reranking reorders candidate chunks using embedding cosine similarity. Under the tested conditions, reranking had limited measurable effect. We interpret this cautiously because the reranking signal is closely related to the retrieval signal, and because small \texttt{top\_k} values and packing behavior can reduce the practical effect of reordering.

\subsection{Packing}

Retrieved chunks are packed into a generation context under an approximate maximum context budget of 452 tokens. The system tracks packed token count, packed chunk count, and whether the gold answer appears in the packed context.

Packing is treated as a separate stage because retrieval success does not guarantee that answer-bearing evidence reaches the generator.

\subsection{Generation}

Generation uses \texttt{google/flan-t5-base} with deterministic decoding and \texttt{max\_new\_tokens = 32} \citep{chung2022flan}. The prompt format is:

\begin{lstlisting}
Context: <packed context>

Question: <question>

Answer briefly using only the context.
If the answer is not present, say NOT_FOUND.
\end{lstlisting}

\subsection{Metrics}

We report final answer metrics together with intermediate diagnostic metrics. Exact Match (EM) is the strict normalized match between prediction and the gold answer. F1 is the token-overlap score between prediction and gold answer. Hit@k indicates whether any retrieved chunk comes from the gold \texttt{context\_id}. Gold-in-packed-context indicates whether the gold answer appears in the final packed context.

These metrics separate stages that are often collapsed into a single score. Hit@k measures whether retrieval reached the correct source context. Gold-in-packed-context measures whether answer-bearing evidence survived packing. EM and F1 measure whether the generator produced the expected answer. Reporting all four allows the analysis to distinguish retrieval failures from packing and generation failures.

We define failure rates as:
\begin{equation}
\text{retrieval\_failure\_rate} =
\frac{\#\{q_i : \text{gold\_chunk\_retrieved}(q_i)=0\}}{N},
\end{equation}
\begin{equation}
\text{packing\_failure\_rate} =
\frac{\#\{q_i : \text{gold\_in\_packed\_context}(q_i)=0\}}{N},
\end{equation}
\begin{equation}
\text{em\_zero\_rate} =
\frac{\#\{q_i : EM(q_i)=0\}}{N}.
\end{equation}

For stability experiments, we compute mean and standard deviation across repeated seeded runs.

\subsection{Experimental Environment}

Experiments were conducted in Google Colab using an NVIDIA T4 GPU runtime with Python 3.10. The implementation used \texttt{transformers} v4.40.x, \texttt{sentence-transformers} v2.7.x, \texttt{faiss-cpu} v1.8.x, and \texttt{torch} v2.2.x. Token-based chunking used the GPT-2 tokenizer, while generation used the FLAN-T5 tokenizer and \texttt{google/flan-t5-base}. Embeddings were normalized before indexing, and a separate \texttt{IndexFlatIP} FAISS index was created for each chunk-size setting to preserve metadata alignment. Generation used greedy decoding with \texttt{max\_new\_tokens = 32} and a 512-token truncation limit.

\section{Experimental Design}

\subsection{Sensitivity Experiments}

The sensitivity study contains 32 baseline runs. We vary chunk size, retrieval depth, and reranking while holding noise at zero:
\begin{align}
\texttt{chunk\_size} &\in \{80,100,120,140\},\\
\texttt{top\_k} &\in \{2,3,4,5\},\\
\texttt{rerank} &\in \{\text{off},\text{on}\}.
\end{align}

The purpose is not to identify a universal best configuration. Instead, the sensitivity study characterizes how different configurations trade off retrieval coverage, context size, latency, and generation quality.

\subsection{Robustness Experiments}

The robustness study contains 12 runs across four anchor configurations (Table~\ref{tab:anchors}). Noise levels are 10\%, 20\%, and 30\%. For each retrieved chunk, noise injection replaces the chunk with a random corpus chunk with probability \texttt{noise\_percent / 100}. The number of retrieved chunks remains fixed. Noise injection and stability experiments use fixed random seeds for reproducibility; Appendix~D provides the pseudocode.

\begin{table}[t]
\centering
\small
\begin{tabular}{lccc}
\toprule
Anchor & Chunk & \texttt{top\_k} & Rerank \\
\midrule
A & 80 & 2 & off \\
B & 80 & 5 & on \\
C & 140 & 2 & on \\
D & 140 & 5 & off \\
\bottomrule
\end{tabular}
\caption{Robustness anchor configurations.}
\label{tab:anchors}
\end{table}

\subsection{Stability Experiments}

The stability study contains 12 repeated-seed runs across four groups (Table~\ref{tab:stability-design}). Each group is repeated with seeds 1, 2, and 3. Noise is fixed at 20\% because it provides a moderate stress condition: strong enough to induce variability, but not so destructive that the pipeline collapses.

\begin{table}[t]
\centering
\small
\begin{tabular}{lcccc}
\toprule
Group & Chunk & \texttt{top\_k} & Rerank & Noise \\
\midrule
A\_80\_k2\_off & 80 & 2 & off & 20 \\
B\_80\_k2\_on & 80 & 2 & on & 20 \\
C\_140\_k5\_off & 140 & 5 & off & 20 \\
D\_140\_k5\_on & 140 & 5 & on & 20 \\
\bottomrule
\end{tabular}
\caption{Stability configurations. Each group is repeated with three seeds.}
\label{tab:stability-design}
\end{table}

\section{Results}

\subsection{Preprocessing Answer Survival}

\begin{figure}[t]
  \centering
  \includegraphics[width=\columnwidth]{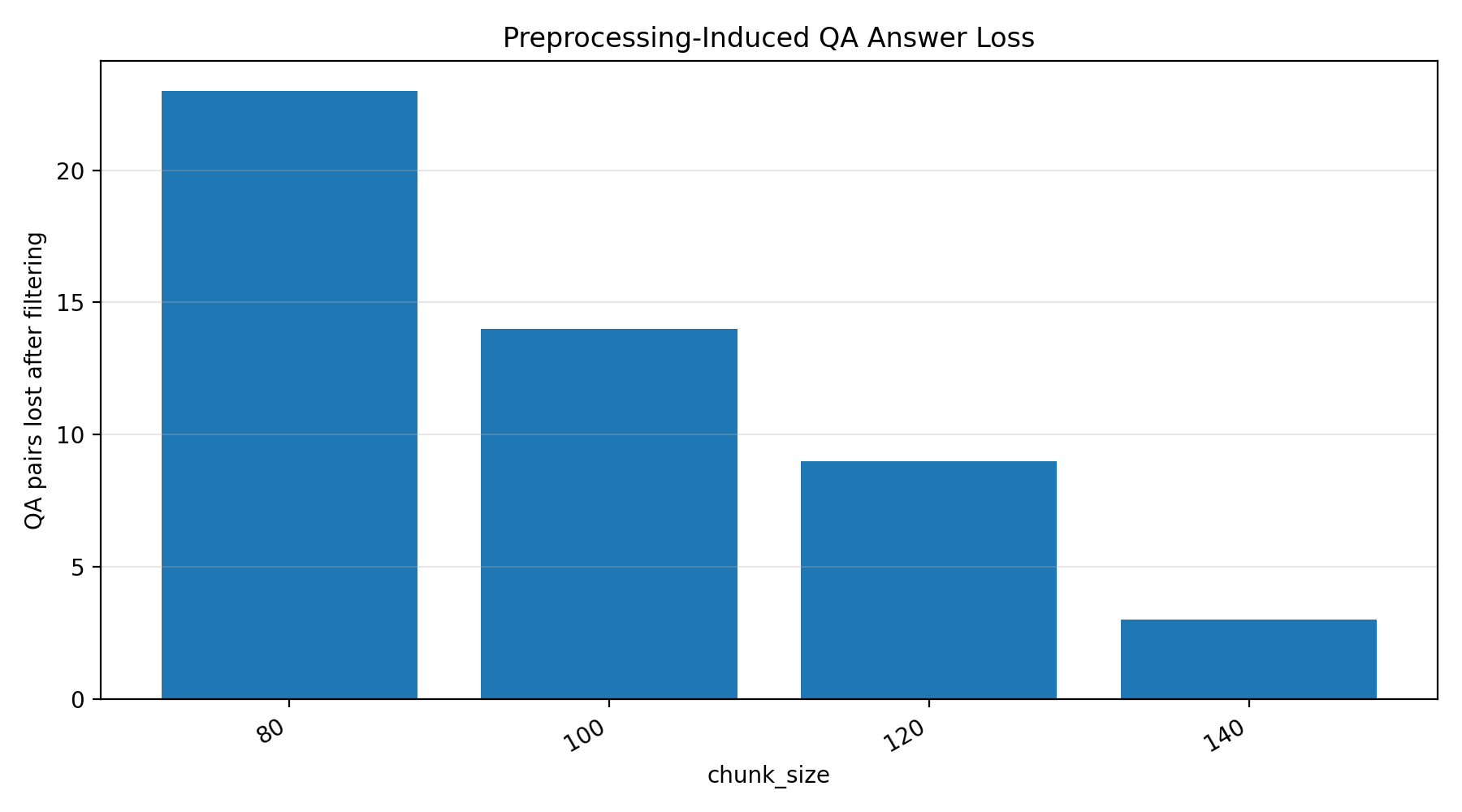}
  \caption{Preprocessing-induced QA answer loss across chunk sizes.}
  \label{fig:preprocess-loss}
\end{figure}

Smaller chunk sizes caused more preprocessing-induced answer loss. With chunk size 80, 23 of the 500 QA pairs lost their answer after filtering. With chunk size 140, only 3 QA pairs were lost (Table~\ref{tab:preprocess-loss}; Figure~\ref{fig:preprocess-loss}).

\begin{table}[t]
\centering
\small
\begin{tabular}{ccc}
\toprule
Chunk size & Preserved & Lost \\
\midrule
80 & 477 & 23 \\
100 & 486 & 14 \\
120 & 491 & 9 \\
140 & 497 & 3 \\
\bottomrule
\end{tabular}
\caption{Preprocessing-induced answer survival by chunk size.}
\label{tab:preprocess-loss}
\end{table}

Smaller non-overlapping chunks fragment contexts more aggressively and produce short tail chunks that are removed by the token-count filter. When an answer appears in a removed chunk or is separated from necessary context, the question becomes unanswerable before retrieval begins.

Chunking is not a neutral preprocessing step. It changes the answerability of the evaluation set and should be considered part of the RAG system behavior. This supports the broader claim that RAG failures can originate before retrieval.

\subsection{Sensitivity to Retrieval Configuration}

\begin{figure}[t]
  \centering
  \includegraphics[width=\columnwidth]{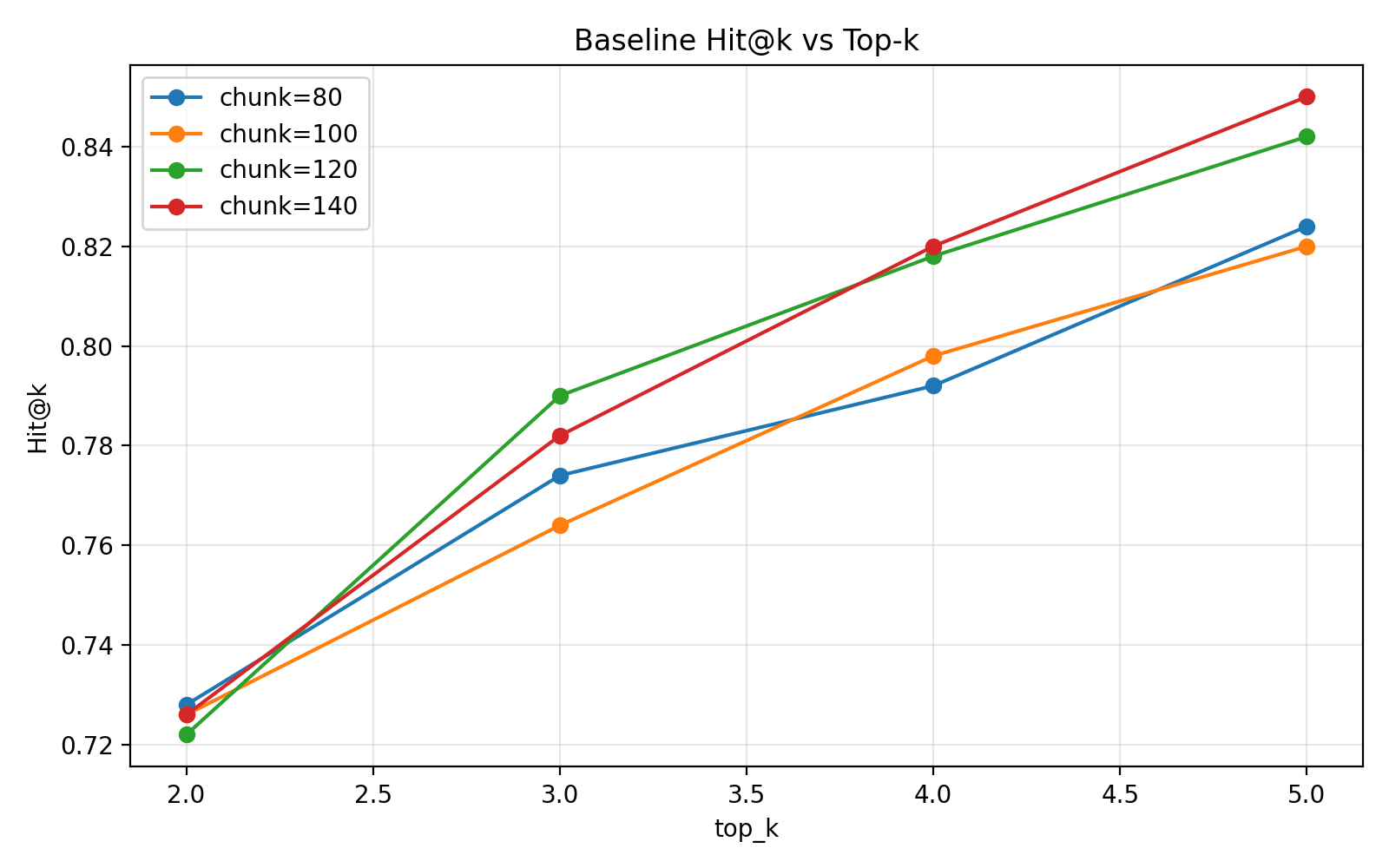}
  \caption{Baseline Hit@k increases with retrieval depth.}
  \label{fig:hit-topk}
\end{figure}

Increasing \texttt{top\_k} improved retrieval coverage. Averaged across baseline configurations, Hit@k increased from 0.726 at \texttt{top\_k=2} to 0.834 at \texttt{top\_k=5}. Gold-in-packed-context also increased from 0.699 to 0.796 across the same range (Figure~\ref{fig:hit-topk}). This indicates that retrieval depth reliably improves coverage, but coverage alone does not determine whether retrieved evidence is usable for generation.

\begin{figure}[t]
  \centering
  \includegraphics[width=\columnwidth]{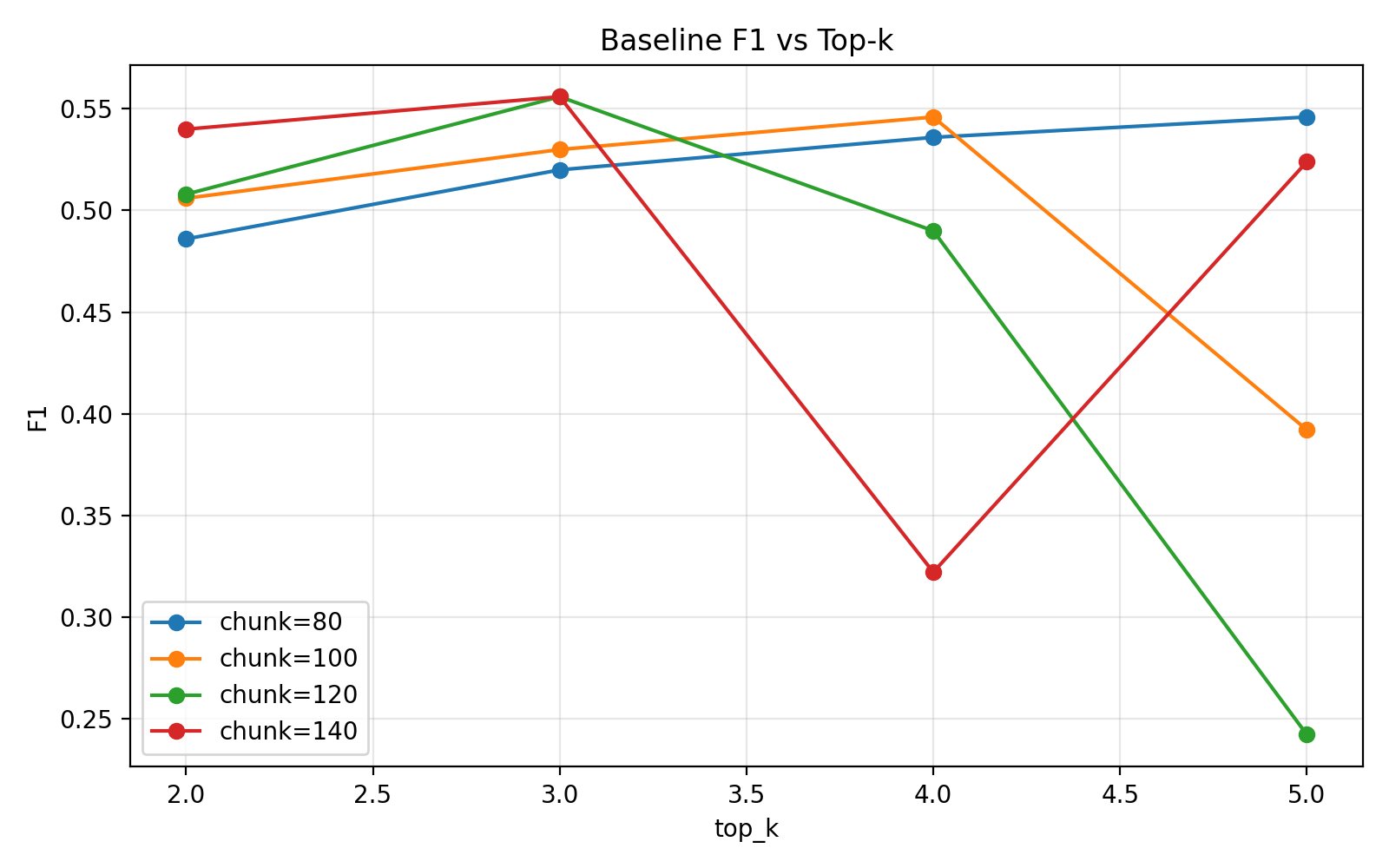}
  \caption{Baseline F1 shows non-monotonic behavior as retrieval depth increases.}
  \label{fig:f1-topk}
\end{figure}

However, final answer quality did not follow the same monotonic pattern. Mean F1 peaked at \texttt{top\_k=3} and then decreased at higher retrieval depths (Table~\ref{tab:topk-means}; Figure~\ref{fig:f1-topk}). Table~\ref{tab:sensitivity-summary} highlights representative configurations that make this tradeoff visible without listing all 32 sensitivity runs.

\begin{table}[t]
\centering
\small
\begin{tabular}{cccccc}
\toprule
\texttt{top\_k} & Hit@k & Gold & EM & F1 & Tokens \\
\midrule
2 & 0.726 & 0.699 & 0.455 & 0.510 & 242.3 \\
3 & 0.778 & 0.749 & 0.481 & 0.540 & 341.3 \\
4 & 0.807 & 0.778 & 0.408 & 0.474 & 425.0 \\
5 & 0.834 & 0.796 & 0.367 & 0.426 & 470.2 \\
\bottomrule
\end{tabular}
\caption{Baseline means grouped by retrieval depth.}
\label{tab:topk-means}
\end{table}

\begin{table*}[t]
\centering
\small
\begin{tabular}{lccccccc}
\toprule
Observation & Chunk & \texttt{top\_k} & Rerank & Hit@k & Gold & EM & F1 \\
\midrule
Strong EM/F1 region & 120 & 3 & off & 0.790 & 0.766 & 0.496 & 0.556 \\
Same region with rerank & 120 & 3 & on & 0.790 & 0.766 & 0.496 & 0.556 \\
Highest Hit@k & 140 & 5 & off & 0.850 & 0.804 & 0.454 & 0.524 \\
Highest gold-in-packed / mismatch & 120 & 5 & off & 0.842 & 0.818 & 0.196 & 0.242 \\
High-context degradation case & 140 & 4 & off & 0.820 & 0.812 & 0.260 & 0.322 \\
\bottomrule
\end{tabular}
\caption{Representative sensitivity configurations. The table emphasizes tradeoffs rather than presenting the full 32-run grid: retrieval-oriented metrics can be strongest in broader regimes, while generation quality can peak elsewhere.}
\label{tab:sensitivity-summary}
\end{table*}

This pattern is one of the clearest examples of non-monotonic RAG behavior in the study. The system retrieves more gold-context evidence at higher \texttt{top\_k}, but the generator does not consistently convert that additional evidence into better answers.

Higher \texttt{top\_k} increases the probability of retrieving the gold context, but it also increases the amount of context passed into generation. Larger packed contexts may dilute attention, introduce irrelevant evidence, or create competition between answer-bearing and distractor content.

\begin{figure}[t]
  \centering
  \includegraphics[width=\columnwidth]{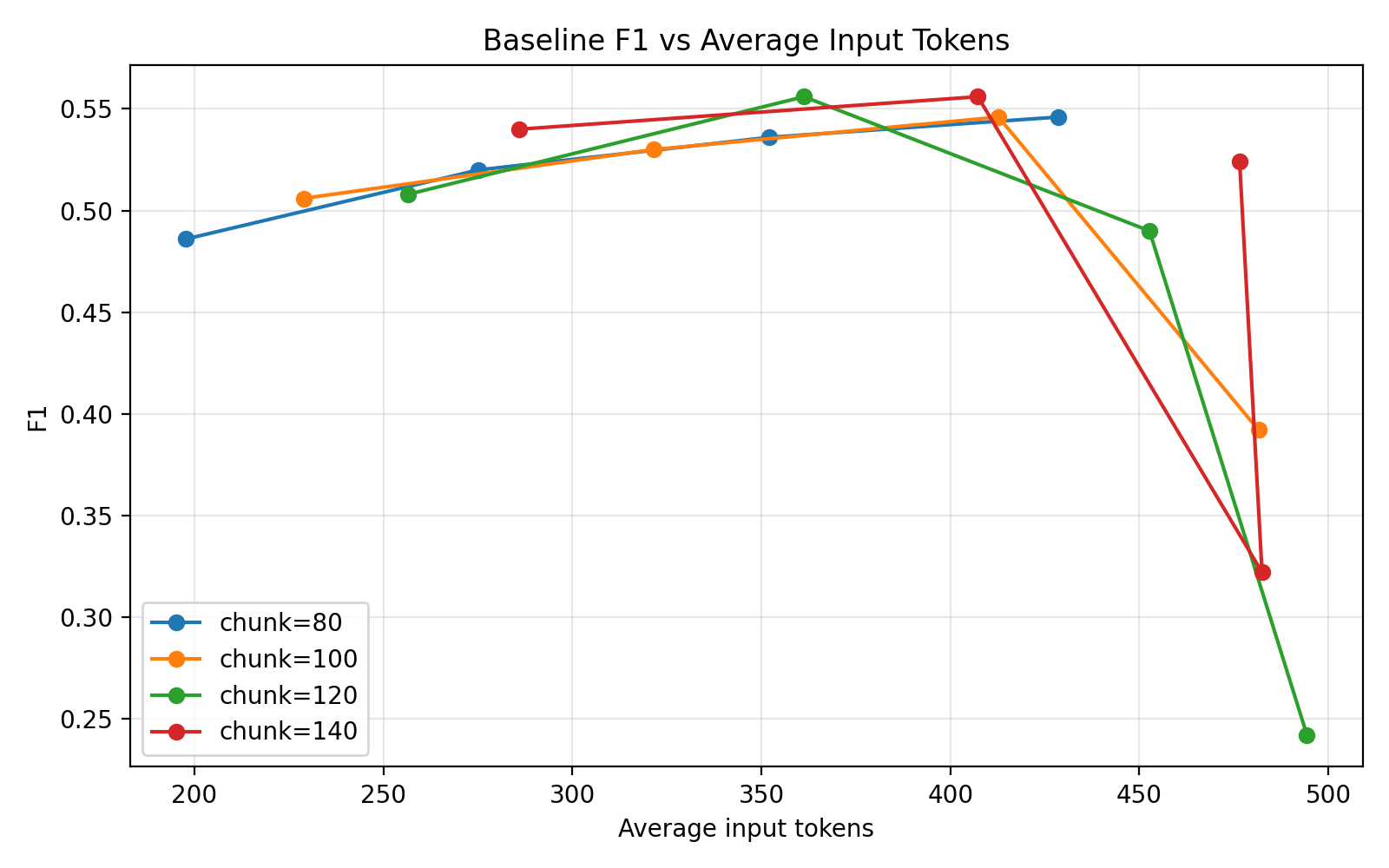}
  \caption{Relationship between input token count and F1, supporting the context-overload interpretation.}
  \label{fig:f1-tokens}
\end{figure}

This is the central retrieval-generation mismatch of the study. Retrieval coverage and packed-context coverage are necessary but not sufficient for answer quality. Figure~\ref{fig:f1-tokens} connects this mismatch to input-token growth: broader retrieval produces longer prompts, and several high-token configurations show weaker F1. A configuration can retrieve more evidence while making generation harder. Therefore, RAG evaluation should report retrieval and generation metrics separately and should avoid treating deeper retrieval as automatically better.

\subsection{Configuration Tradeoffs Rather Than a Universal Optimum}

The strongest EM configuration in the baseline grid was chunk size 120 with \texttt{top\_k=3}, for both rerank off and rerank on. The strongest F1 region included chunk size 120 with \texttt{top\_k=3} and chunk size 140 with \texttt{top\_k=3}. In contrast, the highest Hit@k occurred at chunk size 140 with \texttt{top\_k=5}, while the highest gold-in-packed-context occurred at chunk size 120 with \texttt{top\_k=5}.

The chunk 120, \texttt{top\_k=5} case is especially informative. It achieved the highest gold-in-packed-context rate in the baseline grid, yet its EM and F1 dropped sharply relative to the moderate-depth \texttt{top\_k=3} setting. This is not a retrieval failure in the usual sense; it is a case where evidence availability and generation quality diverge. One plausible explanation is that the broader retrieval depth increased the packed-token load, making the generator more vulnerable to context overload or truncation effects. This interpretation is consistent with the token-growth trend in Table~\ref{tab:topk-means} and Figure~\ref{fig:f1-tokens}. The result makes the retrieval-generation mismatch concrete: placing answer-bearing evidence into the prompt is necessary, but not sufficient, for reliable generation.

This pattern should not be read as a single universal optimum. Instead, different configurations optimize different properties. Broader retrieval improves coverage but increases context length and latency. Moderate retrieval depth can balance coverage and generation quality. Larger chunks reduce preprocessing answer loss but can contribute to heavier contexts. These tradeoffs are the point of the sensitivity analysis.

\subsection{Robustness Under Retrieval Noise}

\begin{figure}[t]
  \centering
  \includegraphics[width=\columnwidth]{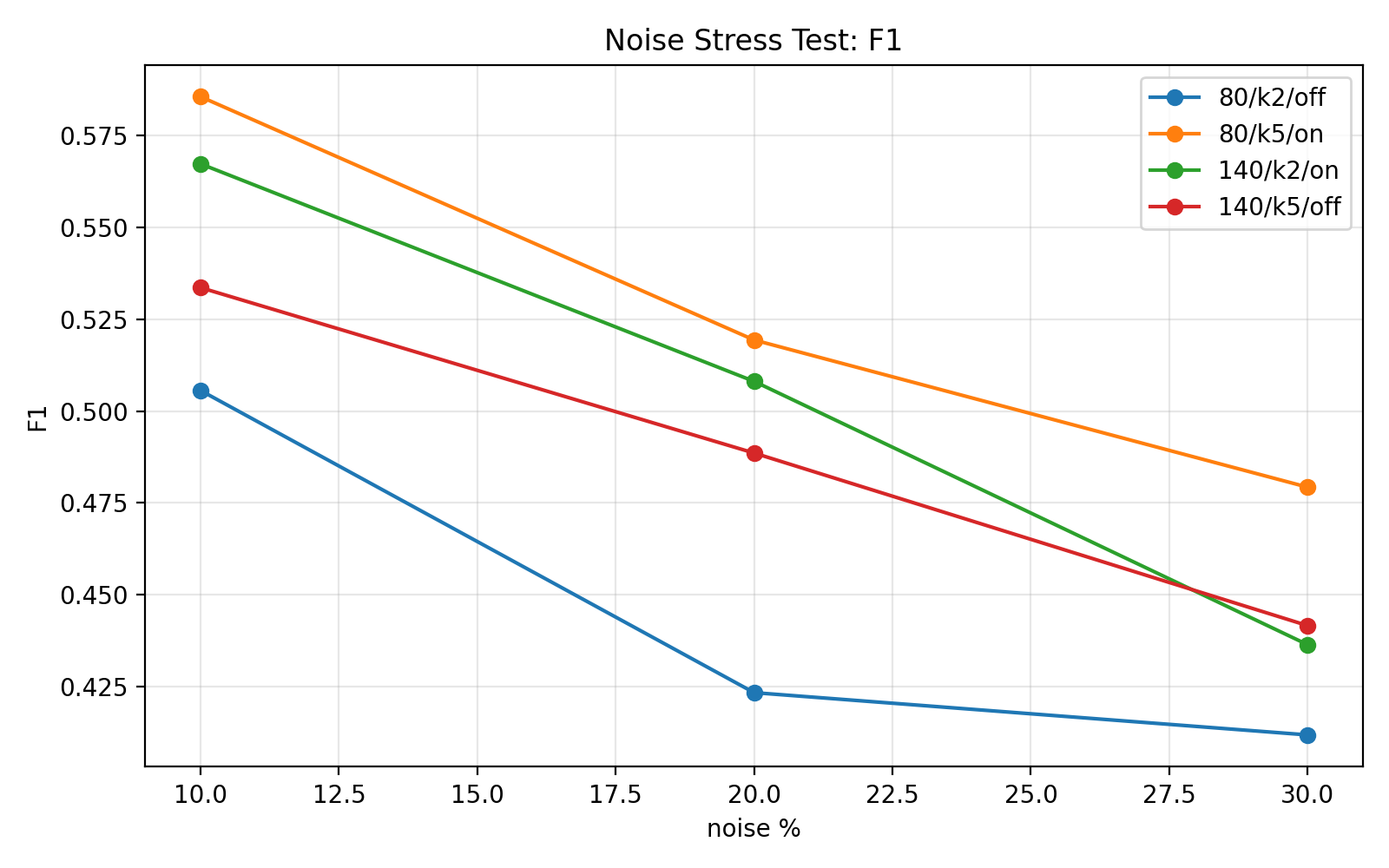}
  \caption{F1 degradation under increasing retrieval corruption.}
  \label{fig:noise-f1}
\end{figure}

\begin{figure}[t]
  \centering
  \includegraphics[width=\columnwidth]{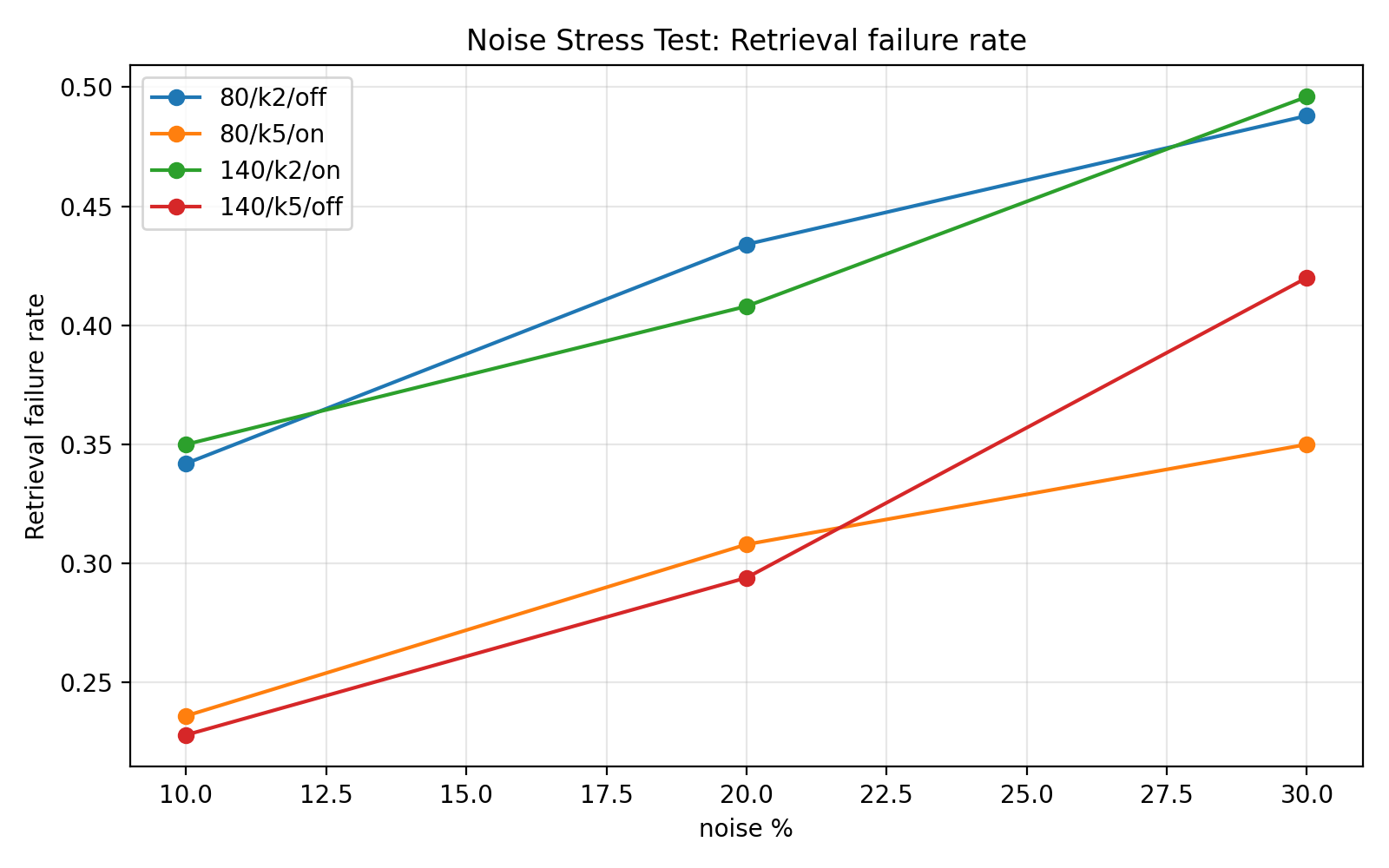}
  \caption{Retrieval failure increases as noise increases.}
  \label{fig:noise-failure}
\end{figure}

Retrieval corruption caused progressive degradation. Averaged across the four robustness anchors, Hit@k decreased from 0.711 at 10\% noise to 0.561 at 30\% noise. Mean F1 decreased from 0.548 to 0.442, while retrieval failure and packing failure increased (Table~\ref{tab:noise-means}; Figures~\ref{fig:noise-f1} and~\ref{fig:noise-failure}). The degradation is gradual rather than abrupt, which suggests that RAG reliability weakens as retrieval evidence is increasingly corrupted instead of failing only after a single threshold.

\begin{table}[t]
\centering
\small
\begin{tabular}{cccccc}
\toprule
Noise & Hit@k & Gold & EM & F1 & Ret. fail \\
\midrule
10\% & 0.711 & 0.672 & 0.421 & 0.548 & 0.289 \\
20\% & 0.639 & 0.604 & 0.373 & 0.485 & 0.361 \\
30\% & 0.561 & 0.541 & 0.339 & 0.442 & 0.439 \\
\bottomrule
\end{tabular}
\caption{Mean robustness metrics by retrieval noise level.}
\label{tab:noise-means}
\end{table}

Random chunk replacement reduces the probability that retrieved evidence comes from the gold context. This affects downstream packing because fewer answer-bearing chunks are available to include in the prompt. Generation quality then declines because the generator receives weaker or missing evidence.

RAG degradation under retrieval corruption is progressive rather than binary. The system does not simply work or fail; instead, errors propagate across retrieval, packing, and generation. This supports the need for robustness evaluation before deployment, especially for applications where retrieval quality may vary.

\subsection{Stability Across Repeated Seeded Runs}

\begin{figure}[t]
  \centering
  \includegraphics[width=\columnwidth]{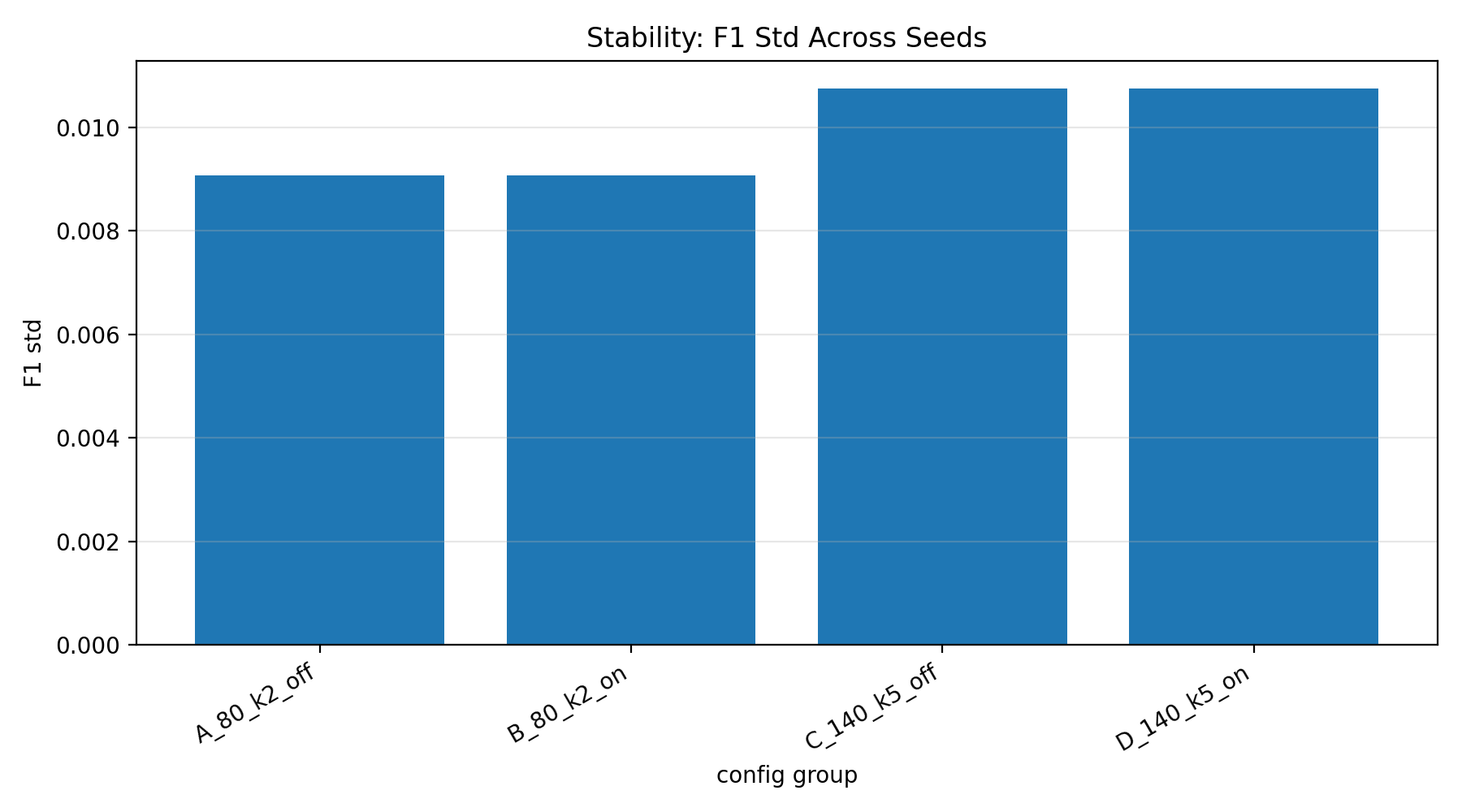}
  \caption{F1 standard deviation across repeated seeded runs.}
  \label{fig:stability-f1}
\end{figure}

The small strict retrieval groups exhibited lower observed variance, while the broader 140/k5 groups showed moderately higher variance. The 80/k2 groups had F1 standard deviation of 0.0091, while the 140/k5 groups had F1 standard deviation of 0.0108 (Table~\ref{tab:stability}; Figure~\ref{fig:stability-f1}).

\begin{table}[t]
\centering
\small
\begin{tabular}{lcccc}
\toprule
Group & EM & EM std & F1 & F1 std \\
\midrule
A\_80\_k2\_off & 0.357 & 0.0058 & 0.456 & 0.0091 \\
B\_80\_k2\_on & 0.357 & 0.0058 & 0.456 & 0.0091 \\
C\_140\_k5\_off & 0.381 & 0.0140 & 0.491 & 0.0108 \\
D\_140\_k5\_on & 0.381 & 0.0140 & 0.491 & 0.0108 \\
\bottomrule
\end{tabular}
\caption{Stability summary across three seeds per group.}
\label{tab:stability}
\end{table}

Broader retrieval regimes include more chunks and therefore more opportunities for random corruption to alter the packed context. Larger chunks and higher \texttt{top\_k} can improve coverage, but they also increase the number and size of context units affected by noise. Because each group uses three seeds, the stability results should be read as observed variance under a controlled stress condition rather than as a broad statistical claim. A lower-variance configuration can still perform poorly, and a higher-performing configuration can show greater variability. This reinforces the paper's tradeoff framing: configurations differ across coverage, generation quality, robustness, latency, and stability.

\subsection{Multi-Stage Failure Analysis}

Failure analysis is one of the core contributions of this study. A final answer failure can arise from at least three distinct stages:
\begin{itemize}
  \item \textbf{Retrieval failure:} no retrieved chunk comes from the gold \texttt{context\_id}.
  \item \textbf{Packing failure:} answer-bearing evidence is not present in the final packed context.
  \item \textbf{Generation failure:} answer-bearing evidence is available, but the generated answer does not match the gold answer.
\end{itemize}

These failure modes should not be collapsed into a single final accuracy metric. For example, an EM-zero result caused by retrieval failure has a different engineering implication from an EM-zero result caused by generation failure after successful retrieval. The first suggests retrieval or indexing issues; the second suggests context use, prompting, or model limitations. This distinction is central to diagnosing RAG behavior.

\section{Discussion}

\subsection{Retrieval-Generation Mismatch}

The main finding is that retrieval and generation quality are not monotonically correlated. Increasing \texttt{top\_k} improves Hit@k and gold-in-packed-context, but answer quality peaks at moderate retrieval depth and declines in broader regimes. One interpretation is that retrieving more evidence can make the generation problem harder, even when it makes the retrieval problem easier.

This mismatch matters because RAG evaluation often reports retrieval-oriented and answer-oriented metrics together, which can make retrieval improvements appear to be a proxy for answer improvements \citep{lewis2020rag,chen2023rgb,saadfalcon2024ares}. Our results support a more cautious interpretation: retrieval coverage is a necessary condition for grounded generation, but the generator must still identify and use the correct evidence inside a limited and potentially noisy context.

\subsection{Context Overload and Attention Dilution}

The sensitivity results support a context-overload interpretation. Higher \texttt{top\_k} increases average input tokens, and higher input-token regimes show weaker EM/F1 in several configurations. This does not isolate a single mechanism, but it is consistent with attention dilution, irrelevant evidence, and conflicting context competing with the answer-bearing passage.

The practical implication is that retrieval depth should be tuned jointly with chunk size and context budget. More context is not automatically better context.

\subsection{Robustness Under Corruption}

Noise experiments show that retrieval corruption causes progressive degradation across the pipeline. The degradation appears in retrieval metrics, packed-context metrics, final answer metrics, and failure rates. This is important for real-world RAG systems because retrieval quality can degrade due to domain shift, ambiguous queries, embedding mismatch, stale indexes, or noisy corpora.

The anchor configurations also show that robustness is configuration-dependent. Broader retrieval can provide redundancy, but it also increases context load. Strict retrieval can be more predictable, but it may be more fragile when the few retrieved chunks are corrupted.

\subsection{Computational Tradeoffs}

The sensitivity results also show a systems cost tradeoff. Increasing retrieval depth and using larger chunks increases the amount of context passed to the generator, which raises input-token usage and can increase latency. Averaged across baseline configurations, increasing \texttt{top\_k} from 2 to 5 raised mean input tokens from 242.3 to 470.2 and mean latency from 1549.9 ms to 2534.1 ms. These costs matter because the same settings that improve retrieval coverage may reduce generation quality or efficiency. In applied RAG systems, configuration tuning should therefore consider retrieval coverage, answer quality, context length, and runtime cost jointly rather than optimizing any single metric in isolation.

\subsection{Stability and Reproducibility}

The stability experiments provide an exploratory view of repeated-run variability under moderate noise. The observed differences are modest, but they support the idea that broader retrieval regimes can show higher variance. Because only three seeds are used per group, these results should be treated as an initial stability analysis rather than a definitive statistical characterization.

\subsection{Reranking Under Tested Conditions}

Reranking had limited measurable effect in both baseline and stability settings. This result should be interpreted carefully. The reranker used an embedding-based similarity signal related to the original retrieval signal, and small \texttt{top\_k} values or packing behavior may have limited the effect of reordering. The result does not imply that reranking is generally ineffective; rather, it indicates that reranking design and signal diversity matter when reranking is expected to improve RAG behavior.

This result highlights that reranking effectiveness depends not only on reranker quality, but also on signal diversity and downstream packing dynamics.

\section{Recommendations}

These results point to practical evaluation habits rather than a single preferred configuration. The recommendations below emphasize separating pipeline stages and tuning for tradeoffs instead of optimizing only final answer accuracy.
\begin{itemize}
  \item Evaluate retrieval and generation separately instead of relying only on EM/F1.
  \item Track gold-in-packed-context and failure rates to identify where failures occur.
  \item Tune \texttt{top\_k} jointly with chunk size and context budget.
  \item Monitor input token length to avoid context overload.
  \item Test retrieval-corruption robustness before deployment.
  \item Measure seed-level stability for critical configurations.
  \item Use stronger or signal-diverse rerankers if reranking is expected to provide measurable gains.
\end{itemize}

\section{Limitations and Threats to Validity}

The current study focuses on controlled experimental settings and leaves broader architectural and retrieval variations for future work. Experiments use a single dense embedding model, MiniLM, and a single generator model, FLAN-T5-base. Results may differ with larger instruction-tuned models, domain-specific retrievers, or stronger generators. Reranking is embedding-based and uses a signal closely related to retrieval; a cross-encoder reranker or hybrid retriever may produce different behavior. Chunking uses non-overlapping token windows only, while overlapping or semantic chunking may reduce answer fragmentation. Retrieval noise is random corruption rather than adversarial or semantically similar distractor injection. Finally, the stability study uses three seeds per group, which is sufficient for exploratory variance analysis but not for broad statistical claims.

Internal validity threats include the use of answer-string matching for gold-in-packed-context and EM/F1, which can understate semantically correct answers or overstate evidence presence when answer strings are ambiguous. We mitigate retrieval ambiguity by using \texttt{context\_id} for Hit@k rather than answer text alone. External validity threats include the fixed 500-question QA subset, the single corpus, and the tested range of retrieval depths.

Results should therefore be interpreted as behavioral observations under the tested pipeline configuration rather than universal properties of all RAG systems.

\section{Conclusion}

This paper presented a controlled empirical analysis of RAG sensitivity, robustness, and stability across 56 runs. The results show that RAG systems are configuration-sensitive and that improved retrieval coverage does not necessarily translate into improved generation quality. Smaller chunks increase preprocessing-induced answer loss, while broader retrieval regimes improve coverage but can introduce context overload and higher variance. Under retrieval corruption, degradation is progressive and propagates through retrieval, packing, and generation. These findings support evaluating RAG systems as multi-stage pipelines rather than only through final answer accuracy. As RAG systems move into production and high-reliability applications, understanding where and why failures emerge may become as important as improving final benchmark accuracy itself.

\section*{Acknowledgements}

The author used generative AI tools, including OpenAI ChatGPT/Codex and Anthropic Claude, for limited assistance with editorial refinement, LaTeX formatting, proofreading, and targeted coding support during development. All experimental design, implementation decisions, analysis, interpretation, and final scientific conclusions were performed and verified by the author.

\bibliographystyle{plainnat}
\bibliography{references}

\appendix

\section{Compact Run Summary}

Table~\ref{tab:appendix-runs} reports representative configurations from the sensitivity and robustness experiments. The selected rows cover the strongest generation region, retrieval-heavy configurations, a retrieval-generation mismatch case, and the 20\% noise anchors. The complete 44-run sensitivity and robustness table is provided in the supplementary material.

\begin{table*}[t]
\centering
\small
\begin{tabular}{lccccccccc}
\toprule
Role & Run & Chunk & k & Rerank & Noise & Hit@k & Gold & EM & F1 \\
\midrule
Best EM / high F1 region & 19 & 120 & 3 & off & 0 & 0.790 & 0.766 & 0.496 & 0.556 \\
Same region with rerank & 20 & 120 & 3 & on & 0 & 0.790 & 0.766 & 0.496 & 0.556 \\
Highest Hit@k & 31 & 140 & 5 & off & 0 & 0.850 & 0.804 & 0.454 & 0.524 \\
Highest gold-in-packed / mismatch case & 23 & 120 & 5 & off & 0 & 0.842 & 0.818 & 0.196 & 0.242 \\
Noise anchor A at 20\% & 37 & 80 & 2 & off & 20 & 0.566 & 0.502 & 0.324 & 0.423 \\
Noise anchor B at 20\% & 38 & 80 & 5 & on & 20 & 0.692 & 0.630 & 0.394 & 0.519 \\
Noise anchor C at 20\% & 39 & 140 & 2 & on & 20 & 0.592 & 0.608 & 0.394 & 0.508 \\
Noise anchor D at 20\% & 40 & 140 & 5 & off & 20 & 0.706 & 0.676 & 0.380 & 0.489 \\
\bottomrule
\end{tabular}
\caption{Representative sensitivity and robustness runs used to connect the appendix to the main-paper tradeoff analysis.}
\label{tab:appendix-runs}
\end{table*}

\section{Stability Group Summary}

Table~\ref{tab:appendix-stability} provides the group-level stability summary used in Section~5.4. The table is kept in the appendix because the main text reports the key mean and variance patterns directly.

\begin{table*}[t]
\centering
\small
\begin{tabular}{lccccccccc}
\toprule
Group & Chunk & k & Rerank & Noise & Seeds & EM & EM std & F1 & F1 std \\
\midrule
A\_80\_k2\_off & 80 & 2 & off & 20 & 3 & 0.357 & 0.0058 & 0.456 & 0.0091 \\
B\_80\_k2\_on & 80 & 2 & on & 20 & 3 & 0.357 & 0.0058 & 0.456 & 0.0091 \\
C\_140\_k5\_off & 140 & 5 & off & 20 & 3 & 0.381 & 0.0140 & 0.491 & 0.0108 \\
D\_140\_k5\_on & 140 & 5 & on & 20 & 3 & 0.381 & 0.0140 & 0.491 & 0.0108 \\
\bottomrule
\end{tabular}
\caption{Stability group summary from \texttt{stability\_master\_group\_summary.csv}.}
\label{tab:appendix-stability}
\end{table*}

\section{Prompt}

\begin{lstlisting}
Context: <packed context>

Question: <question>

Answer briefly using only the context.
If the answer is not present, say NOT_FOUND.
\end{lstlisting}

\section{Noise Injection Pseudocode}

\begin{lstlisting}
for chunk in retrieved_chunks:
    if random() < noise_percent / 100:
        chunk = random_chunk_from_corpus()
\end{lstlisting}

\section{Graph Catalog}

Main paper figures include the pipeline architecture diagram, preprocessing answer loss, baseline Hit@k vs. \texttt{top\_k}, baseline F1 vs. \texttt{top\_k}, baseline F1 vs. input tokens, noise F1, noise retrieval failure, and stability F1 standard deviation. Additional plots are available in the supplementary material, including EM vs. \texttt{top\_k}, gold-in-packed-context vs. \texttt{top\_k}, latency vs. \texttt{top\_k}, noise Hit@k, noise EM, noise gold-in-packed-context, noise packing failure, noise EM-zero failure, average noisy chunks, stability EM standard deviation, stability Hit@k standard deviation, and stability F1 mean.

\end{document}